\documentclass[12pt]{article}
\pdfoutput=1
\usepackage[pdftex,hidelinks]{hyperref}
\usepackage[pdftex]{graphicx}
\usepackage[pdftex]{color}
\usepackage{mathtools,amssymb,amsmath}
\usepackage{booktabs}
\usepackage{physics}
\usepackage{subfigure}
\usepackage{float}
\usepackage{slashed}
\usepackage{multirow}
\usepackage{arydshln}
\usepackage[centering]{geometry}
\usepackage{cite}

\allowdisplaybreaks[4]

\makeatletter
\@addtoreset{equation}{section}
\makeatother

\begin{document}
\setlength{\baselineskip}{0.6cm}

\thispagestyle{empty}

\begin{flushright}
NITEP-189
\end{flushright}
\vspace{40pt}
\begin{center}

{\Large\bf{Toward Realistic Models \\
\vspace*{3mm}
in $T^2/\mathbb{Z}_2$ Flux Compactification}} \\

\vspace{40pt}

{\bf{Hiroki Imai$^\dagger \hspace{.5pt}$}\footnote{E-mail: sw23928v@st.omu.ac.jp}}, \,
{\bf{Nobuhito Maru$^{\dagger, \, \ast}\hspace{.5pt}$}}

\vspace{40pt}
{\it $^{\dagger}$ Department of Physics, 
        Osaka Metropolitan University, \\
        Osaka 558-8585, Japan \\[5pt]
$^{\ast}$ Nambu Yoichiro Institute of Theoretical and Experimental Physics (NITEP), \\
        Osaka Metropolitan University, Osaka 558-8585, Japan} \\
\end{center}
\vspace{30pt}
\begin{abstract}
	\noindent
We consider a six dimensional gauge theory compactified on $T^2/\mathbb{Z}_2$ with magnetic flux. The configurations of models are classified by winding numbers at the fixed points. Requiring the existence of generation numbers and Yukawa coupling, we see that allowed and forbidden configurations are described by geometry of winding numbers.
\end{abstract}

\newpage
\setcounter{page}{2}
\setcounter{footnote}{0}

\section{Introduction}
\label{sec:1}
The Standard Model (SM) of particle physics has achieved remarkable successes, but there remain some mysteries. We see that there are three-generation of quarks and leptons by observation, but the origin is not revealed and there might be a fourth generation discovered in future experiments. We know that the masses of quarks/leptons are generated by the Yukawa couplings with Higgs boson, but these coupling constants with large hierarchy are given by hand and the origin is unknown. Therefore, we have to consider physics beyond the Standard Model (BSM).

Higher dimensional theory with compactified extra dimension is one of the candidates of BSM. We can consider a scenario in which a single fermion in higher dimension has a degeneracy with respect to four-dimensional (4d) quarks/leptons and the degree of degeneracy determines the generation number. This is achieved by introducing a magnetic flux in extra dimensions\cite{WITTEN:1984, GSW, CIM:2004}. SM fermions are chiral zero modes and their existence is supported by the Atiyah-Singer (AS) index theorem\cite{Atiyah:1968}.

Furthermore, we also consider orbifolding of the extra dimension. Magnetized orbifolds might provide rich flavor structures due to orbidold fixed points\cite{Abe:2008, Abe:2009, Fujimoto:2013, Abe:2014_ZN_twisted, Abe:2014_Operator_analysis}. In $T^2/\mathbb{Z}_N \, (N=2,3,4,6)$, winding numbers at the fixed points play an important role to classify configurations of models and contribute the generation number\cite{Sakamoto:2020, Sakamoto:2021, Imai:2022}.\footnote{In this context, some studies consider magnetized blow-up manifolds \cite{Kobayashi:2019_blow-up, Kobayashi:2020_blow-up, Kobayashi:2023_blow-up}. In our analysis, we do not consider to remove the singularities like these.}

In general, higher dimensional models have many configurations. Some configurations are realistic and others are not. Only realistic configurations are candidates of BSM. Indeed, classification between realistic and unrealistic configurations of magnetized $T^2/\mathbb{Z}_N \, (N=2,3,4,6)$ is studied in Refs.\cite{Abe:2015_Classification, Fujimoto:2016}. However, the geometrical meanings behind these results are still unclear. 

In this paper, we clarify that the geometrical meaning is given by the winding numbers at the fixed points in magnetized $T^2/\mathbb{Z}_2$. We require some conditions from phenomenological point of view, which are the existence of generations and Yukawa coupling. The former implies the presence of multi-generation of fermions and at least one ``generation'' of the Higgs boson in 4d physics. The latter is needed to realize fermion masses. 

By imposing these conditions, we can obtain the possible number of the SM Higgs boson. We start from 6d Weyl fermions and consider two cases where 4d left- and right-handed fermions are originated from the 6d Weyl fermions with same 6d chirality or different one. We cannot give any constraint in the former case, which has many Higgs bosons. In the latter case, however, we see that certain geometric patterns of the winding numbers does not give positive Higgs number, which must be excluded. We note that the result does not depend on specific generation numbers and Yukawa coupling constants, thus we obtain general properties of realistic models in a magnetized orbifold.

This paper is organized as follows. 
In Section \ref{sec:2}, we review 6d models compactified on $T^2/\mathbb{Z}_2$ with magnetic flux and see that configurations of models are determined by winding numbers at the fixed points.
In Section \ref{sec:3}, we require the existance of the generations and Yukawa coupling in magnetized $T^2/\mathbb{Z}_2$ models to constrain configurations by a geometry of winding numbers at fixed points. 
Section \ref{sec:4} is devoted to conclusion and discussion. 
\section{Set up}
\label{sec:2}
In this section, we briefly review 6d models compactified on $T^2$ or $T^2/\mathbb{Z}_2$ 
with magnetic flux \cite{Imai:2022, Sakamoto:2020, Abe:2014_ZN_twisted, Abe:2014_Operator_analysis}. 
We emphasize that the generation number of the SM fermions is given by the index theorem.
\subsection{$T^2$ model}
\label{subsec:2-1}
\begin{figure}[t]
    \centering
    \includegraphics[height=5cm]{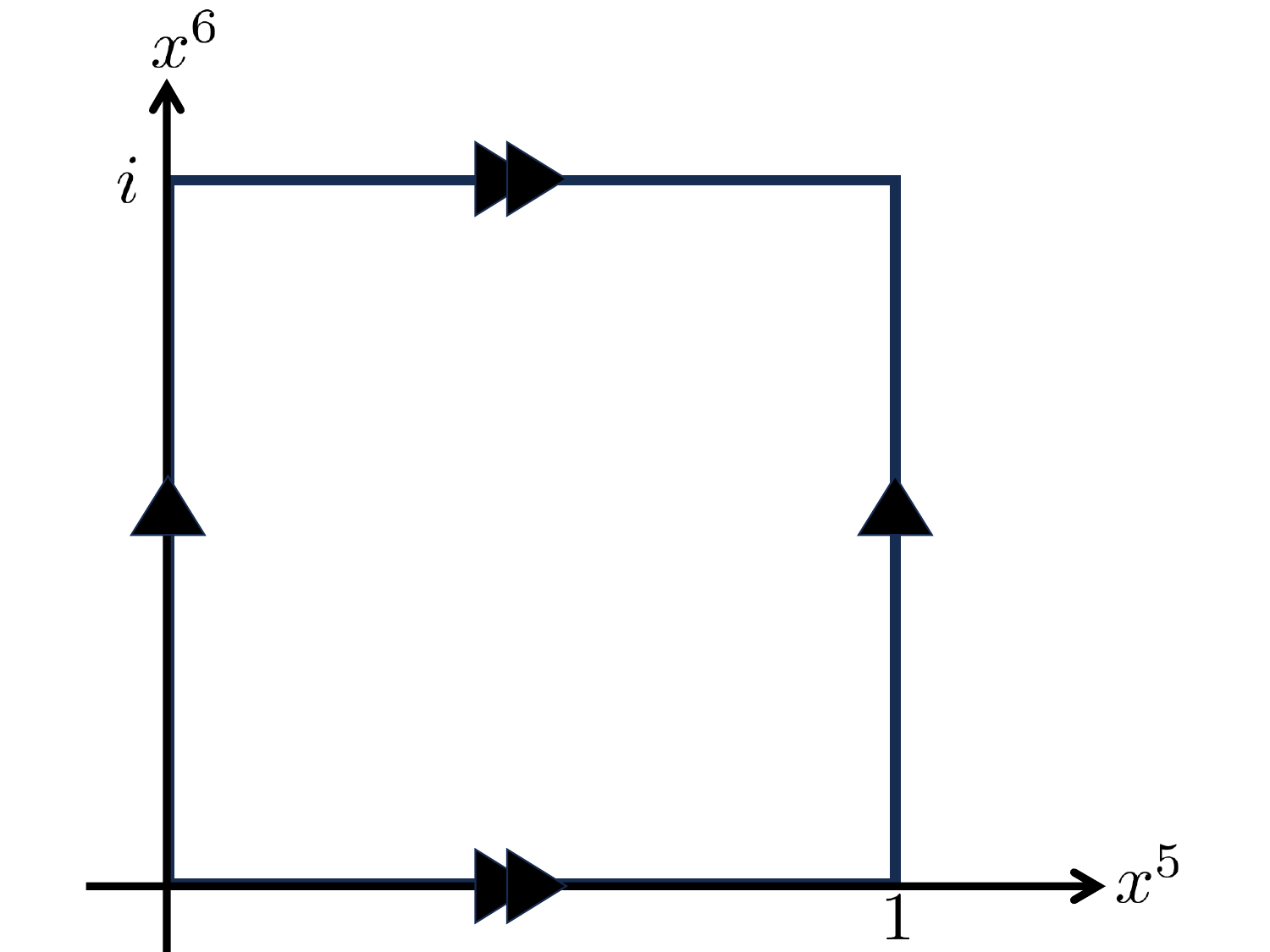}
    \caption{The torus is formed by identifying along the directions of the arrows.}
    \label{fig:2-1}%
\end{figure}

First of all, we consider a 6d spacetime where the extra dimensions are compactified on a two dimensional (2d) torus $T^2$.
By use of the complex coordinate $z=x^5+ix^6$, 
the torus $T^2$ is defined by two identifications between $x^5=0$ and $x^5=1$, and between $x^6=0$ and $x^6=1$ (See Fig. \ref{fig:2-1})
\footnote{For simplicity, we take the radii of the torus to be 1 and torus moduli to be $i$.}, i.e.
\begin{equation}
z \sim z+1\sim z+i.
\label{2-1}%
\end{equation}
We introduce a homogeneous static magnetic field $B$ perpendicular to the $x^5-x^6$ plane.
The vector potential is given by
\begin{equation}
	A_5(z)= -\frac12 B x^6=\frac{iB}{4}(z-\bar{z}), \qquad 
	A_{6}(z)= \frac12 Bx^5=\frac{B}{4}(z+\bar{z}),
\label{2-2}%
\end{equation}
which can be rewritten as
\begin{equation}
	A_z(z)\coloneqq \frac12(A_5-iA_6)= -\frac{i}{4}B \bar{z}, \qquad 
	A_{\bar{z}}(z)\coloneqq \frac12(A_5+iA_6)= \frac{i}{4}Bz.
\label{2-3}%
\end{equation}
The torus lattice shifts on the vector potential should be accompanied by the gauge transformations as
\begin{align}
	A_{z/\bar z}(z+1) & = A_{z/\bar z} (z) +\partial_{z/\bar z} \Lambda_1(z),
	\label{2-4}\\
	A_{z/\bar z}(z+i) & = A_{z/\bar z} (z) +\partial_{z/\bar z} \Lambda_2(z),
	\label{2-5}%
\end{align}
where $\Lambda_1(z)$ and $\Lambda_2(z)$ are gauge transformation parameters given by
\begin{align}
	\Lambda_1(z) = \frac12 B \mathrm{Im} z
	,\qquad
	\Lambda_2(z) = -\frac12 B \mathrm{Im}(i z).
\label{2-6}%
\end{align}

We next consider some 6d Weyl fermions in the magnetic flux background. 
The Lagrangian reads
\begin{equation}
	\mathcal{L}_{\text{6d}}= i \sum_f \bar \Psi^f_{\chi^f} \Gamma^K D_K \Psi^f_{\chi^f}
	,\qquad \Gamma_7 \Psi^f_{\chi^f}=\chi^f \Psi^f_{\chi^f},
\label{2-7}%
\end{equation}
where $K \, (=0,1,2,3,5,6)$ is the 6d spacetime index and $f$ is the label specifying the SM fermions. 
In Section \ref{sec:3}, $f$ is used to distinguish between left-handed fermions L and right-handed fermions R.
$D_K=\partial_K - iq^f A_K$ is the covariant derivative, where $q^f$ is the $U(1)$ charge of the fermion. 
$\Gamma^K \, (K=0,1,2,3,5,6)$ denote 6d Gamma matrices and $\Gamma_7=-\Gamma^0\Gamma^1\Gamma^2\Gamma^3\Gamma^5\Gamma^6$ is the 6d chiral operator. 
We use the chiral representation for Gamma matrices\footnote{$\sigma^a \, (a=1,2,3)$ are the Pauli matrices and $I_N$ is the $N\times N$ unit matrix.}: 
 \begin{gather}
 \begin{split}
	\Gamma^\mu = \gamma^\mu\otimes I_2 = 
	\begin{pmatrix}
		\gamma^\mu & 0 \\
		0 & \gamma^\mu
	\end{pmatrix} \qquad (\mu = 0, 1, 2, 3), \\
	\Gamma^5 = i\gamma_5\otimes \sigma_1=
	\begin{pmatrix}
		0 & i\gamma_5 \\
		i\gamma_5 & 0
	\end{pmatrix}, \qquad 
	\Gamma^6 = i\gamma_5\otimes \sigma_2=
	\begin{pmatrix}
		0 & \gamma_5 \\
		-\gamma_5 & 0
	\end{pmatrix}.
\end{split}
\label{gamma-1}
\end{gather}
The representation diagonalize $\Gamma_7$ such as
\begin{gather}
\Gamma_7 = \gamma_5\otimes\sigma_3=
	\begin{pmatrix}
		\gamma_5 & 0 \\
		0 & - \gamma_5
	\end{pmatrix}.
\label{gamma-2}
\end{gather}
We take each fermion $\Psi^f_{\chi^f}$ an eigenstate of $\Gamma_7$, whose eigenvalue is $\chi^f=\pm1$.
$\Psi^f_{\chi^f}$ are called 6d Weyl fermions, 
which are obtained by acting the projection operators $(I_8+\chi^f\Gamma_7)/2$ on 6d Dirac fermions $\Psi^f$ 
such as
\begin{align}
\Psi^f_{\chi^f}&= \frac{I_8+\chi^f \Gamma_7}2 \Psi^f. 
	\label{2-8}
\end{align}
6d Weyl fermions are expanded in terms of Kaluza-Klein (KK) modes,  
\begin{align}
\Psi^f_+(x,z)&=\sum_{n=0}^\infty \left(\psi^f_{\text{4d,R},n}(x)\otimes \psi^f_{\text{2d,}+,n}(z)+\psi^f_{\text{4d,L},n}(x)\otimes \psi^f_{\text{2d,}-,n}(z)\right),
	\label{2-9}\\
\Psi^f_-(x,z)&=\sum_{n=0}^\infty \left(\psi^f_{\text{4d,R},n}(x)\otimes \psi^f_{\text{2d,}-,n}(z)+\psi^f_{\text{4d,L},n}(x)\otimes \psi^f_{\text{2d,}+,n}(z)\right),
	\label{2-10}
\end{align}
which can be written by use of $\chi^f$ as
\begin{equation}
\Psi^f_{\chi^f}(x,z)=\sum_{n=0}^\infty \left(\psi^f_{\text{4d,R},n}(x)\otimes \psi^f_{\text{2d,}\, \chi^f \! ,n}(z)+\psi^f_{\text{4d,L},n}(x)\otimes \psi^f_{\text{2d,}\, -\chi^f \! ,n}(z)\right),
	\label{2-10-1-1}
\end{equation}
where $n$ means the label of KK mass eigenstates (i.e. Landau levels), and $x$ denotes a 4d Minkowski coordinate. 
$\psi^f_{\text{4d,R/L},n}(x)$ are 4d right/left-handed fermions 
which satisfy $\gamma_5 \psi^f_{\text{4d,R/L},n}(x)=\pm\psi^f_{\text{4d,R/L},n}(x)$, 
and $\psi^f_{\text{2d},\pm,n}(z)$ are 2d Weyl spinors 
which satisfy $\sigma_3\psi^f_{\text{2d},\pm,n}(z)=\pm\psi^f_{\text{2d},\pm,n}(z)$. 
The eigenvalue $\pm$ of $\sigma_3$ is called the 2d chirality.

For convenience, we use the following notation:
\begin{gather}
	\psi^f_{\text{2d},+,n} (z)= 
	\begin{pmatrix}
		\psi^f_{T^2,+, n} (z)\\[3pt] 0
	\end{pmatrix}
	, \qquad \psi^{f}_{\text{2d},-, n}(z)= 
	\begin{pmatrix}
		0 \\[3pt] \psi^f_{T^2,-, n}(z),
	\end{pmatrix}.
	\label{2-11}%
\end{gather}
where $\psi^{f}_{T^2,\pm, n}(z)$ are mode functions on $T^2$ with 2d chirality $\pm$. 
The 2d Weyl fermions in general obey the twisted periodic boundary conditions 
associated with the gauge transformation:
\begin{align}
\psi^f_{T^2,\pm, n} (z+1) &= U_1^f(z)\psi^f_{T^2,\pm, n} (z),
\label{2-10-1}\\
\psi^f_{T^2,\pm, n} (z+i) &= U_2^f(z)\psi^f_{T^2,\pm, n} (z)
\label{2-10-2}
\end{align}
with
\begin{equation}
U_j^f(z) = e^{iq^f \Lambda_j(z)}e^{2i\pi \alpha_j^f}\qquad(j=1,2),
\label{2-SS}
\end{equation}
where $(\alpha^f_1,\alpha^f_2)$ is called Scherk-Shwarz twist phase (SS phase).

Due to the consistency of the boundary conditions (\ref{2-10-1}) and (\ref{2-10-2}),  
the magnetic flux has to be quantized as
\begin{equation}
\frac{q^f B}{2\pi} \eqqcolon M^f \in \mathbb{Z},
\label{2-quanta}
\end{equation}
where we call the flux quanta for the quantized magnetic flux number $M^f$. We should emphasize that the flux quanta $M^f$ depends on $q^f$, therefore the fields feel different magnetic flux if the charge is different.

We require that the mode functions on $T^2$ with $\chi^f =+1$ obey the equations
\begin{equation}
\begin{cases}
\slashed D \psi^f_{\text{2d,}+,n}(z)=m_n\psi^f_{\text{2d,}-,n}(z),\\
\slashed D \psi^f_{\text{2d,}-,n}(z)=-m_n\psi^f_{\text{2d,}+,n}(z),
\end{cases}
\label{2-13}
\end{equation}
and those with $\chi^f = -1$ obey the equations
\begin{equation}
\begin{cases}
\slashed D \psi^f_{\text{2d,}+,n}(z)=-m_n\psi^f_{\text{2d,}-,n}(z),\\
\slashed D \psi^f_{\text{2d,}-,n}(z)=m_n\psi^f_{\text{2d,}+,n}(z),
\end{cases}
\label{2-14}
\end{equation}
where the Dirac operator is defined as $\slashed D \coloneqq \sigma_1 D_5 +\sigma_2 D_6$.
Eqs.(\ref{2-13}) and (\ref{2-14}) are Dirac equations on the extra dimension, 
which leads to 4d Dirac equations
\begin{equation}
\begin{cases}
i\gamma^\mu D_\mu \psi^f_{\text{4d,R},n}(z)=m_n\psi^f_{\text{4d,L},n}(z),\\
i\gamma^\mu D_\mu \psi^f_{\text{4d,L},n}(z)=m_n\psi^f_{\text{4d,R},n}(z).
\end{cases}
\label{2-15}
\end{equation}

The SM fermions are given by zero modes of Eqs.(\ref{2-13}) or (\ref{2-14}):
\begin{equation}
\begin{cases}
\slashed D \psi^f_{\text{2d,}+,0}(z)=0,\\
\slashed D \psi^f_{\text{2d,}-,0}(z)=0.
\end{cases}
\label{2-16}
\end{equation}
It follows from Eq.(\ref{2-16}) that the zero mode functions $\psi_{T^2,\pm,0}^f(z)$ on $T^2$ obey
\begin{align}
D_{\bar z}\psi_{T^2,+,0}^f(z)=\left(\partial_{\bar z} +\frac{\pi M^f}{2} z \right)\psi_{T^2,+,0}^f(z)=0, \label{2-16-1} \\
D_{z}\psi_{T^2,-,0}^f(z)=\left(\partial_{ z} -\frac{\pi M^f}{2} \bar{z} \right)\psi_{T^2,-,0}^f(z)=0. \label{2-16-2}
\end{align}
We should note that only Eq.(\ref{2-16-1}) has solutions if $M^f>0$, and only Eq.(\ref{2-16-2}) has those if $M^f<0$\footnote{We can confirm it by the explicit calculation of the mode function which is given by the theta function\cite{CIM:2004}. }. Each zero mode solution is in general degenerate as the lowest-lying mode of the Landau level.

The generation number of the SM fermions is given by the degree of degeneracy of the solutions of Eq.(\ref{2-16-1}) or Eq.(\ref{2-16-2}), which is given by
\begin{equation}
n^f = |M^f|. 
\label{2-17-1}
\end{equation}
This can be understood it from the viewpoint of the AS index theorem. Let $n_\pm^f$ be the numbers of zero modes with 2d chirality $\pm$, which are numbers of solutions of Eqs.(\ref{2-16}). We define the analytical index of the Dirac operator $i\slashed{D}$ as $\mathrm{Ind}(i\slashed{D})\coloneqq n^f_+ - n^f_-$. From the AS index theorem, the index is equal to the flux quanta $M^f$ on the magnetized torus\cite{WITTEN:1984, GSW}:
\begin{equation}
\mathrm{Ind}(i\slashed D) = M^f.
\label{2-17}
\end{equation}
Although the left-hand side of Eq.(\ref{2-17}) is the difference between $n^f_+$ and $n^f_-$, it is consistent with (\ref{2-17-1}) since $n_-^f$ is zero if $M^f>0$ and $n_+^f$ is zero if $M^f<0$. This discussion tells us that $M^f= + 3(M^f=-3)$ gives three-generation of right-(left-)handed fermions with $\chi^f=+1$ and left-(right-)handed fermions with $\chi^f=-1$, respectively.

\subsection{$T^2/\mathbb{Z}_2$ model}
\label{subsec:2-2}

\begin{figure}[t]
    \centering
    \includegraphics[height=5cm]{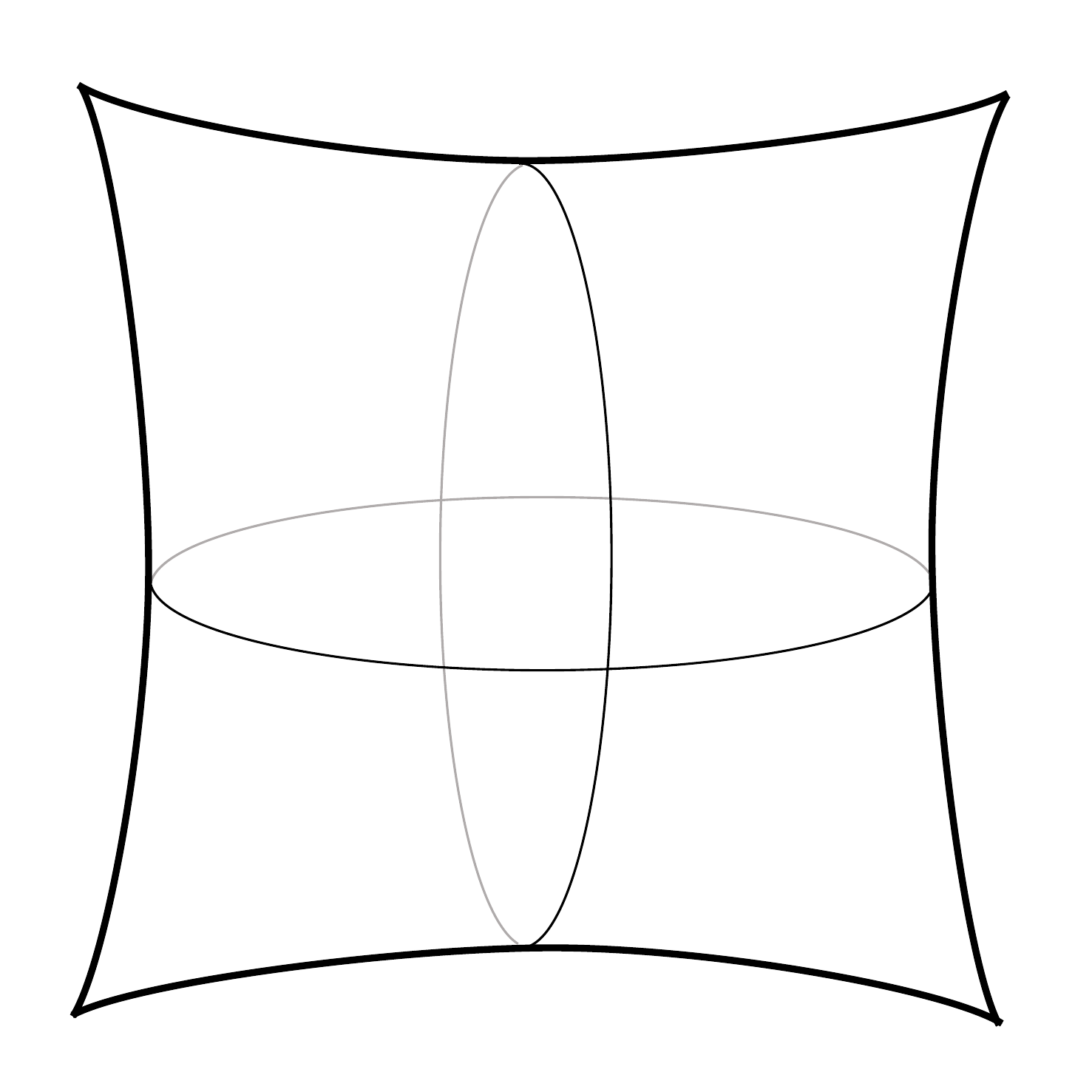}
    \caption{An orbifold $T^2/\mathbb{Z}_2$ forms a pillowcase shape.}
    \label{fig:2-2}%
\end{figure}

Let us consider an orbifolding in this subsection.
An orbifold $T^2/\mathbb{Z}_2$ is defined by the torus identification Eq.(\ref{2-1}) 
and an additional $\mathbb{Z}_2$ parity one
\begin{equation}
z\sim -z,
\label{2-18}
\end{equation}
i.e. the orbifold is obtained from the identification
\begin{equation}
\forall m,n \in \mathbb{Z}, \qquad 
z\sim -z +m +in.
\label{2-19}
\end{equation}
The identification (\ref{2-19}) forms a pillowcase shape illustrated in Fig. \ref{fig:2-2}. 
The important feature is the existence of fixed points $z^{\text{fp}}$ defined by
\begin{equation}
\exists m,n \in \mathbb{Z}, \qquad z^{\text{fp}}= -z^{\text{fp}} +m +in.
\label{2-20}
\end{equation}
There are four fixed points $z^{\text{fp}}_I \, (I=1,2,3,4)$ on the $T^2/\mathbb{Z}_2$ orbifold as
\begin{equation}
	z_1^{\text{fp}}=0, \quad
	z_2^{\text{fp}}=1/2, \quad
	z_3^{\text{fp}}=i/2, \quad
	z_4^{\text{fp}}=(1+i)/2,
\label{2-21}%
\end{equation}
which are corners of the pillowcase.

Now, we consider mode functions on the $T^2/\mathbb{Z}_2$. 
Similar to those on $T^2$, we use the notation:
\begin{gather}
	\psi^f_{\text{2d},+,n} (z)= 
	\begin{pmatrix}
		\psi^f_{T^2/\mathbb{Z}_2,+, n} (z)_{\eta^f}\\[3pt] 0
	\end{pmatrix}
	, \qquad \psi^{f}_{\text{2d},-, n}(z)= 
	\begin{pmatrix}
		0 \\[3pt] \psi^f_{T^2/\mathbb{Z}_2,-, n}(z)_{\eta^f}
	\end{pmatrix},
	\label{2-22}%
\end{gather}
where $\psi^f_{T^2/\mathbb{Z}_2,\pm, n}(z)_{\eta^f}$ are mode functions on $T^2/\mathbb{Z}_2$ 
with $\mathbb{Z}_2$ eigenvalue $\pm\eta^f$:
\begin{align}
\psi^f_{T^2/\mathbb{Z}_2,+, n} (-z)_{\eta^f} &= \eta^f\psi^f_{T^2/\mathbb{Z}_2,+, n} (z)_{\eta^f} ,
\label{2-23-1}\\
\psi^f_{T^2/\mathbb{Z}_2,-, n} (-z)_{\eta^f} &= -\eta^f \psi^f_{T^2/\mathbb{Z}_2,-, n} (z)_{\eta^f} .
\label{2-23-2}
\end{align}
Eqs.(\ref{2-23-1}) and (\ref{2-23-2}) represent the parity transformation on the extra dimension. 
We emphasize that if $\mathbb{Z}_2$ eigenvalue of $\psi^f_{T^2/\mathbb{Z}_2,+, n} (z)_{\eta^f}$ is $\eta^f$, 
then that of  $\psi^f_{T^2/\mathbb{Z}_2,-, n} (z)_{\eta^f}$ has to be  $-\eta^f$. 
We should note that we write the $\mathbb{Z}_2$ eigenvalue of the mode function belonging to 2d chirality $+$ as the subscript at the bottom right. As stated in Ref.\cite{Abe:2014_ZN_twisted}, 6d Weyl fermions are transformed under $\mathbb{Z}_2$ parity as
\begin{equation}
\Psi^f_{\chi^f}(x,-z) = \eta^f \mathcal{S}\Psi^f_{\chi^f}(x,z),
\label{2-23-3} 
\end{equation}
where $\mathcal{S}=\mathrm{diag}(I_4,-I_4)$.

Since $\psi^f_{T^2/\mathbb{Z}_2,\pm, n} (z)_{\eta^f}$ are mode functions on $T^2$ 
as well as those on the $T^2/\mathbb{Z}_2$,
they follow boundary conditions (\ref{2-10-1}) and (\ref{2-10-2}).
From these conditions, the SS phase $(\alpha^f_1,\alpha^f_2)$ has to be quantized as 
\begin{equation}
	(\alpha^f_1,\alpha^f_2) = (0,0), \, (1/2,0), \, (0,1/2), \, (1/2, 1/2).
\label{2-23}%
\end{equation}

The winding numbers at the fixed points give the generation number of fermions. It is given by Ref.\cite{Sakamoto:2020}
\begin{equation}
n^f = \frac{|M^f|}{2} - \frac{V^f}{2}+1,
\label{2-27}
\end{equation}
where $V^f$ is sum of $\mathbb{Z}_2$ winding numbers at the four fixed points.

The winding numbers at the fixed points $\rho^f_{I} \, (I=1,2,3,4)$  are defined by
\begin{equation}
\lim_{z\to0} \psi^f_{T^2/\mathbb{Z}_2,\pm, n} (z^{\text{fp}}_I - z)_{\eta^f}
=
(-1)^{\rho^f_{I}}\lim_{z\to0} \psi^f_{T^2/\mathbb{Z}_2,\pm, n} (z^{\text{fp}}_I + z)_{\eta^f}.
\label{2-25}
\end{equation}
Since the definition (\ref{2-25}) has mod 2 ambiguities of $\rho^f_{I}$,
we consider $\rho^f_{I}$ as either 0 or $1$ when we apply to Eq.(\ref{2-27}).

As in the case of $T^2$, $n^f$ is the number of zero modes,
which is the degree of degeneracy of the solutions of Eq.(\ref{2-16-1}) or Eq.(\ref{2-16-2}) under the conditions (\ref{2-10-1}), (\ref{2-10-2}) and (\ref{2-23-1}), (\ref{2-23-2}).
Eq.(\ref{2-27}) can be understood from the viewpoint of the index theorem on $T^2/\mathbb{Z}_2$ orbifold\cite{Imai:2022}. The index theorem is given by
\begin{equation}
\mathrm{Ind}(i\slashed{D}) = \frac{M^f}{2} + \frac{-V^f_+ + V^f_-}{4},
\label{2-24}
\end{equation}
where $V_\pm^f$ are sums of the winding numbers at the fixed points with 2d chirality $\pm$.
Due to $V_+^f + V_-^f=4$\cite{Imai:2022,Sakamoto:2021},
Eq.(\ref{2-24}) can be also transformed as
\begin{equation}
\mathrm{Ind}(i\slashed D)
=\frac{M^f}{2} - \frac{V^f_+}{2}+1
=\frac{M^f}{2} + \frac{V^f_-}{2}-1,
\label{2-26}
\end{equation}
which is consistent with Eq.(\ref{2-27}).
Since $n^f_+(n^f_-)$ vanishes for $M^f>0(M^f<0)$, we see that the former(latter) equality in Eq.(\ref{2-26}) is reduced to Eq.(\ref{2-27}), respectively.
We can see it from the first equal sign if $M^f>0$ then $n_-^f$ is zero,
and from the second equal sign if $M^f<0$ then $n_+^f$ is zero.

Eq.(\ref{2-27}) can be also applied to the counting of the zero modes for scalar fields although it is not supported by index theorem. In this paper, we simply assume that Eq.(\ref{2-27}) to be correct for the SM Higgs field\footnote{We are assuming that the SM Higgs field is a zero mode of some kind of 4d scalar field, which may be an extra component of higher dimensional gauge field or that in supersymmetric theories. In any case, we can apply Eq.(\ref{2-27}) for Higgs field as long as it is a zero mode.}.

\begin{figure}[h]
    \centering
    \includegraphics[height=15cm]{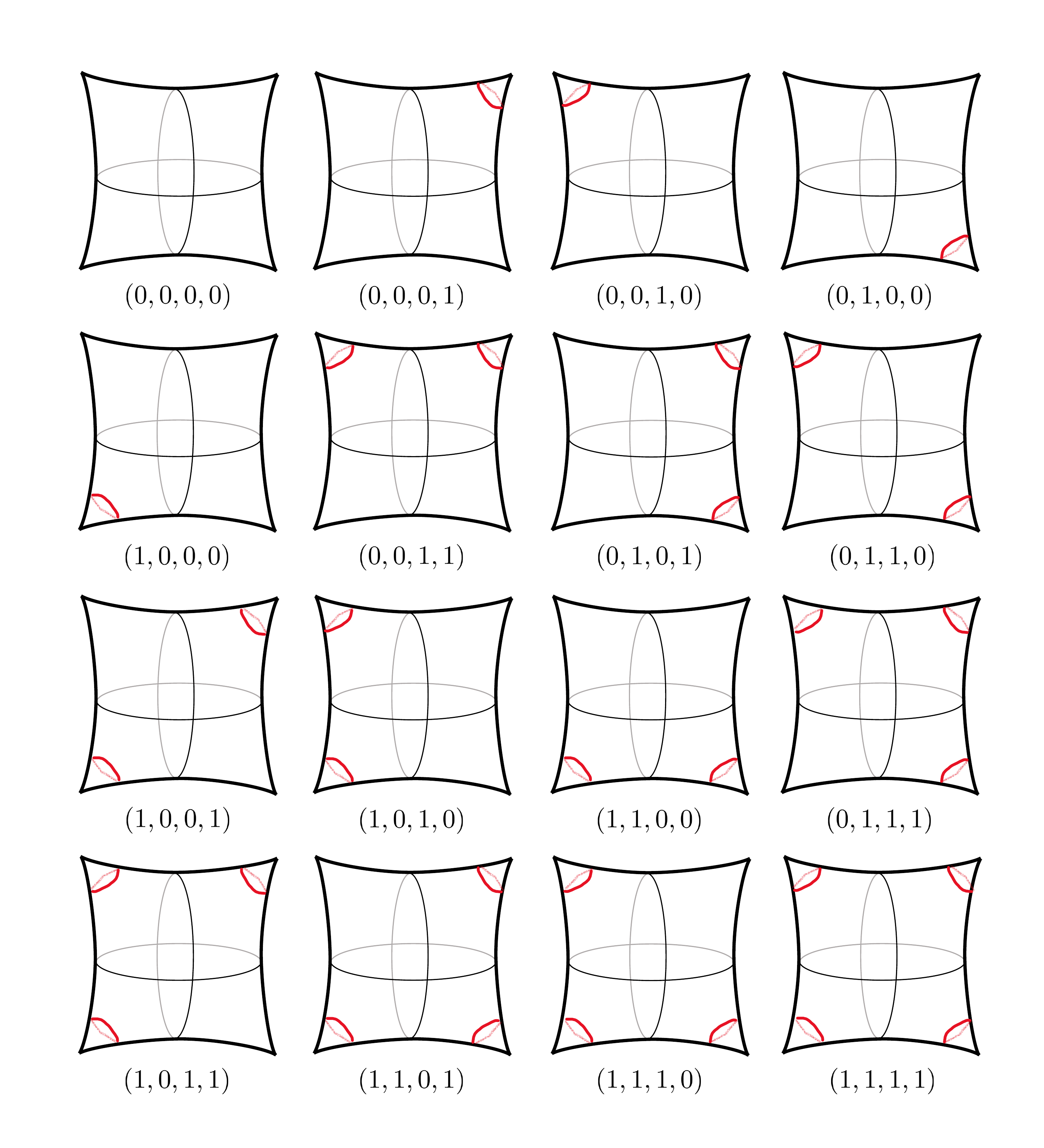}
    \caption{Classification of patterns in a magnetized $T^2/\mathbb{Z}_2$ model are shown schematically. The red lines are the windings of the field around the fixed points. Each picture corresponds to the configuration specified by $(\rho_1^f,\rho_2^f,\rho_3^f,\rho_4^f)$.}
    \label{fig:classification-1}%
\end{figure}

There are $2^4=16$ patterns of models with a magnetized $T^2/\mathbb{Z}_2$ from possible combinations of the winding numbers $(\rho_1^f,\rho_2^f,\rho_3^f,\rho_4^f)$. In Ref.\cite{Abe:2014_ZN_twisted}, the patterns are completely classified by the flux quanta $|M^f|$, the SS phase $(\alpha^f_1,\alpha^f_2)$ and the $\mathbb{Z}_2$ eigenvalue $\eta^f =\pm1$. This classification is equivalent to that of winding numbers as shown in TABLE V of Ref.\cite{Sakamoto:2020}. We rewrite it in Table \ref{table:1} to make it clear to understand the classification by winding numbers. We notice that generation number $n^f$ is determined by the sum of the winding numbers $V^f=0,1,2,3,4$. We also illustrate possible patterns of winding numbers schematically in Fig.\ref{fig:classification-1}.
\begin{table}[h]
\centering
	\begin{tabular}{c|cccc|cccc}
	\hline \hline
	\addlinespace[2pt]
	$V^f$&$\rho_1^f$&$\rho_2^f$&$\rho_3^f$&$\rho_4^f$& $n^f$&$|M^f|$&$(\alpha_1^f,\alpha_2^f)$&$\eta^f$ \\
	\hline
	\addlinespace[2pt]
	0&0&0&0&0&$\displaystyle\frac{|M^f|}2 +1$&even&$(0,0)$&$+1$ \\
	\addlinespace[2pt]
	\hline
	\multirow{4}{*}{1}&0&0&0&1&\multirow{4}{*}{$\displaystyle\frac{|M^f|+1}2$}&\multirow{4}{*}{odd}&$(0,0)$&$+1$\\
	&0&0&1&0&&&$\left(0,\frac12\right)$&$+1$ \\[2pt]
	&0&1&0&0&&&$\left(\frac12,0\right)$&$+1$ \\[2pt]
	&1&0&0&0&&&$\left(\frac12,\frac12\right)$&$-1$ \\[2pt]
	\hline
	\multirow{6}{*}{2}&0&0&1&1&\multirow{6}{*}{$\displaystyle\frac{|M^f|}2$}&\multirow{6}{*}{even}
	&$\left(0,\frac12\right)$&$+1$\\[2pt]
	&0&1&0&1&&&$\left(\frac12,0\right)$&$+1$ \\[2pt]
	&0&1&1&0&&&$\left(\frac12,\frac12\right)$&$+1$ \\[2pt]
	&1&0&0&1&&&$\left(\frac12,\frac12\right)$&$-1$ \\[2pt]
	&1&0&1&0&&&$\left(\frac12,0\right)$&$-1$ \\[2pt]
	&1&1&0&0&&&$\left(0,\frac12\right)$&$-1$ \\[2pt]
	\hline
	\multirow{4}{*}{3}&0&1&1&1&\multirow{4}{*}{$\displaystyle\frac{|M^f|-1}2$}
	&\multirow{4}{*}{odd}&$\left(\frac12,\frac12\right)$&$+1$\\[2pt]
	&1&0&1&1&&&$\left(\frac12,0\right)$&$-1$ \\[2pt]
	&1&1&0&1&&&$\left(0,\frac12\right)$&$-1$ \\[2pt]
	&1&1&1&0&&&$(0,0)$&$-1$ \\
	\hline
	\addlinespace[2pt]
	4&1&1&1&1&$\displaystyle\frac{|M^f|}2 -1$&even&$(0,0)$&$-1$ \\
	\addlinespace[2pt]
	\hline \hline
	\end{tabular}
	\caption{Classification of patterns in a magnetized $T^2/\mathbb{Z}_2$ model. 
	$n_f$ is determined by $V^f$.}
	\label{table:1}
	\end{table}
\clearpage

\section{Toward realistic models}
\label{sec:3}
In this section, we give some requirements to realize generation numbers and fermion masses, and see what kind of configurations satisfy these requirements. The geometrical meaning of the realistic configurations is clarified.

\subsection{Realistic conditions}
\label{subsec:3-1}
In this subsection, we impose the following three conditions in our analysis. 
\begin{itemize}
\item[(i)]The generation numbers of the left- and right-handed fermions are equal.
\item[(ii)]At least one SM Higgs boson exists.
\item[(iii)]Yukawa coupling should be allowed.
\end{itemize}
These requirements are necessary for realistic model building.

The condition (i) and (ii) claim the existence of generation numbers. The former means that the generation number does not depend on chirality although the left- and right-handed fermions are independent in the SM. The latter means that the ``generation'' number of Higgs is more than one, which is clear from the experimental fact.

The condition (iii) is required to give masses to quarks/leptons.
In this paper, we consider Yukawa coupling in the 6d bulk\footnote{
There is another set up that Yukawa coupling is localized at the fixed points\cite{Ishida:2017}.}:
\begin{equation}
\mathcal{L}_{\text{Yukawa}}
=
\bar\Psi^{\text{L}}_{\chi^{\text{L}}} H \Psi^{\text{R}}_{\chi^{\text{R}}} + \text{h.c.},
\label{Yukawa}
\end{equation}
where $\Psi^{\text{L}}_{\chi^{\text{L}}}$ and $\Psi^{\text{R}}_{\chi^{\text{R}}}$ are 6d Weyl fermions generating the left- and right-handed 4d fermions in the SM, respectively. $H$ is a 6d scalar\footnote{In the following \ref{subsec:3-3}, we consider the both cases where $H$ is an extra dimensional component of a gauge field
(Case I) and where it is just a 6d scalar field (Case II).} field including the SM Higgs boson as the zero mode.

Note that specific values of the generation numbers and Yukawa coupling constants are not assumed in our analysis, which is applicable to the cases where the fourth generation of quarks/leptons or multi-Higgs doublets are present.

\subsection{Formulations of the conditions}
\label{subsec:3-2}
In this subsection, the above three conditions are discussed in more detail.

To express the conditions (i) and (ii), we use the generation formula Eq.(\ref{2-27}).
Denoting the number of the left-(right-)handed massless fermion as $n^{\text{L}}$($n^{\text{R}}$), the condition (i) means $n^{\text{L}}=n^{\text{R}} \eqqcolon g$, which can be expressed by using Eq.(\ref{2-27}),
\begin{equation}
\frac{|M^{\text{L}}|}{2}-\frac{V^{\text{L}}}{2} +1
=
\frac{|M^{\text{R}}|}{2}-\frac{V^{\text{R}}}{2} +1=g
\label{constraint-0}
\end{equation}
and is simplified as
\begin{equation}
|M^{\text{L}}|-V^{\text{L}} = |M^{\text{R}}| - V^{\text{R}}=2g-2.
\label{constraint-1}
\end{equation}
The condition (ii), which realizes at least one ``generation'' number of Higgs, is
\begin{equation}
n^{\text{H}}=
\frac{|M^{\text{H}}|}{2}-\frac{V^{\text{H}}}{2} +1\geq 1,
\label{3-8}
\end{equation}
which is reduced to
\begin{equation}
|M^{\text{H}}|\geq V^{\text{H}}.
\label{3-9}
\end{equation}
Note that we do not need specific values of $g$ and $n^{\text{H}}$ in our analysis.\\

To express the condition (iii), we have to mention boudary conditions. For the Yukawa coupling (\ref{Yukawa}) to be consistent with the boundary conditions from magnetized $T^2/\mathbb{Z}_2$, it must be invariant under the torus lattice shifts (\ref{2-10-1}), (\ref{2-10-2}) and $\mathbb{Z}_2$ parity transformation (\ref{2-23-1}), (\ref{2-23-2}).

Let us first consider the transformation under the torus lattice shifts. Since the phases appearing in the Yukawa coupling must be vanished,
\begin{align}
&-M^{\text{L}} + M^{\text{H}} +M^{\text{R}} =0, \label{constraint-2-1}\\
&-\alpha^{\text{L}}_1 + \alpha^{\text{H}}_1 + \alpha^{\text{R}}_1 = 0, \label{constraint-2-2} \\
&-\alpha^{\text{L}}_2 + \alpha^{\text{H}}_2 + \alpha^{\text{R}}_2 = 0. \label{constraint-2-3} 
\end{align}
are required\cite{Abe:2015_Classification}. We consider that the field $H$ feels the magnetic field whose flux quanta\footnote{Since we consider that the field $H$ has the $U(1)$ charge $q^{\text{H}}$, it can feel magnetic flux different from fermions.} is $M^{\text{H}}$ and satisfied twisted boudary conditions with the SS phase $(\alpha_1^{\text{H}},\alpha_2^{\text{H}})$.
These Higgs parameters are completely determined by the left- and right-handed fermion parameters from Eqs.(\ref{constraint-2-1})-(\ref{constraint-2-3}).

Next we consider the transformation under $\mathbb{Z}_2$ parity. Eq.(\ref{2-23-3}) tells us that the first term of Eq.(\ref{Yukawa}) transforms as
\begin{equation}
\overline{\left(\eta^{\text{L}}\mathcal{S}\Psi^{\text{L}}_{\chi^{\text{L}}}\right)}\eta^{\text{H}}H\left(\eta^{\text{R}}\mathcal{S}\Psi^{\text{R}}_{\chi^{\text{R}}}\right)
=
\bar \eta^{\text{L}} \eta^{\text{H}}\eta^{\text{R}}
\bar\Psi^{\text{L}}_{\chi^{\text{L}}} \Gamma^0\mathcal{S}\Gamma^0 H \mathcal{S}\Psi^{\text{R}}_{\chi^{\text{R}}},
\label{con-1}
\end{equation}
where we used $\left(\Gamma^0\right)^2=I_8$ and $\mathcal{S}^\dagger =\mathcal{S}$. 
If $H$ is an extra component of higher dimensional field, it is written as linear combination of $\Gamma^5$ and $\Gamma^6$. In this case, since we find $\left[\mathcal{S},\Gamma^0\right]=0$ and $\left\{\mathcal{S},\Gamma^5\right\}=\left\{\mathcal{S},\Gamma^6\right\}=0$, 
\begin{equation}
\bar \eta^{\text{L}} \eta^{\text{H}}\eta^{\text{R}}=-1
\label{con-2}
\end{equation}
is required for the Yukawa coupling to be allowed.
On the other hand, if $H$ is just a scalar,
\begin{equation}
\bar \eta^{\text{L}} \eta^{\text{H}}\eta^{\text{R}}=1
\label{con-3}
\end{equation}
is required. Due to Lorentz symmetry in 6d theory, $\chi^{\text{L}}=\chi^{\text{R}}$ is required in the former case and $\chi^{\text{L}}=-\chi^{\text{R}}$ is required in the latter case. After all, 
\begin{equation}
-\chi^{\text{L}}\chi^{\text{R}}  \bar\eta^{\text{L}} \eta^{\text{H}}\eta^{\text{R}} = 1. \label{constraint-2-4}
\end{equation}
is required for the Yukawa coupling to be allowed. 

These conditions (\ref{constraint-2-1})-(\ref{constraint-2-3}) and (\ref{constraint-2-4}) are rewritten by the winding numbers as
\begin{equation}
\frac{\chi^{\text{L}}+\chi^{\text{R}}}{2} - \rho^{\text{L}}_I +\rho^{\text{H}}_I +\rho^{\text{R}}_I = 0 \mod 2 \qquad \text{for } \forall I\in\{1,2,3,4\}.
\label{constraint-2} 
\end{equation}
The detail check of Eq.(\ref{constraint-2}) is given in Appendix \ref{Appendix-1}.

\subsection{Analysis toward realistic models}
\label{subsec:3-3}
In this subsection, we perform analysis by depending on two ways of chirality assignments.
As a result, we obtain a geometrical property to constrain the configurations in one case.
\subsubsection*{Case I : $\chi^{\text{\rm{L}}}=\chi^{\text{\rm{R}}}$}
In this case, 4d left- and right-handed fermions come from the 6d Weyl fermion with same chirality\footnote{In this case, we assume that the SM Higgs field is originated from extra components of higher dimensional field and Yukawa coupling is given by gauge coupling.}, $\chi^{\text{L}}=\chi^{\text{R}}=+1$ or $-1$. 
Eq.(\ref{constraint-2}) is written as
\begin{equation}
\forall I \in \{1,2,3,4\}, \qquad
1-\rho^{\text{L}}_I + \rho^{\text{R}}_I + \rho^{\text{H}}_I =0 \mod 2.
\label{3-10-0}
\end{equation}
To eliminate mod 2 ambiguity, we take $\rho^{\text{L,R,H}}$ to be 0 or 1. 
Then,
\begin{equation}
\forall I \in \{1,2,3,4\}, \qquad
\rho_I^{\text{H}} = 1 - |-\rho_I^{\text{L}} + \rho_I^{\text{R}}|.
\label{3-10}
\end{equation}
Since the left- and the right-handed quarks/leptons have the same 6d chirality, 
their 2d chiralities are opposite each other. 
This means that the magnetic flux $M^{\text{L}}$ and $M^{\text{R}}$ have opposite signs. 
For instance, in the case of $\chi^{\text{L}}=\chi^{\text{R}}=+1$, 
which is the same set up in Ref.\cite{Hoshiya:2022}, $M^{\text{L}}<0$ and $M^{\text{R}}>0$ are required. 
The relation between flux quanta (\ref{constraint-2-1}) tells us
\begin{equation}
M^{\text{H}} = M^{\text{L}} - M^{\text{R}} =
\begin{cases}
 -|M^{\text{L}}| -|M^{\text{R}}|&\text{for }\chi^{\text{L}}=\chi^{\text{R}}=1, \\
 |M^{\text{L}}|+|M^{\text{R}}|&\text{for }\chi^{\text{L}}=\chi^{\text{R}}=-1,
\end{cases}
\label{3-11}
\end{equation}
therefore we see $|M^{\text{H}}|=|M^{\text{L}}|+|M^{\text{R}}|$ in either case. 
The number of Higgs fields is given by
\begin{align}
n^{\text{H}} &= \frac{|M^{\text{H}}|}{2} -\frac{V^{\text{H}}}{2} +1 \nonumber \\
&= \frac{|M^{\text{L}}|+|M^{\text{R}}|}{2} -\frac12\sum_{I=1}^4 \left(1- |-\rho_I^{\text{L}} 
+ \rho_I^{\text{R}}| \right)+1 \nonumber \\
&=\frac{|M^{\text{L}}|+|M^{\text{R}}|}{2}+\frac12 \sum_{I=1}^4  |-\rho_I^{\text{L}} + \rho_I^{\text{R}}|  -1,
\label{3-12}
\end{align}
where we use Eq.(\ref{3-10}) in the second equality. Since we find $|M^{\text{L}}|=V^{\text{L}}+2g-2$ and $|M^{\text{R}}|=V^{\text{R}}+2g-2$ from Eq.(\ref{constraint-1}),
the number of Higgs fields can be expressed in term of the generation number $g$ as
\begin{align}
n^{\text{H}} &= 2g+ \frac{V^{\text{L}}+V^{\text{R}}}{2}+\frac12 
\sum_{I=1}^4  |-\rho_I^{\text{L}} + \rho_I^{\text{R}}|  -3 \nonumber \\
&=2g-3 +\frac12\sum_{I=1}^4(\rho_I^{\text{L}} + \rho_I^{\text{R}}+|-\rho_I^{\text{L}} + \rho_I^{\text{R}}|  ). 
\label{3-14-1} 
\end{align}
Since $\rho_I^{\text{L/R}} $ are 0 or 1, $\rho_I^{\text{L}} + \rho_I^{\text{R}}+|-\rho_I^{\text{L}} + \rho_I^{\text{R}}| =0,2$ for all $I \in \left\{1,2,3,4\right\}$.
Therefore the last term of Eq.(\ref{3-14-1}) takes an integer in the range $[0,4]$, and we obtain 
\begin{equation}
2g-3 \leq n^{\text{H}}  \leq 2g+1
\label{3-14}
\end{equation}
In three generation case $g=3$, the number of Higgs fields is more than 3 and less than 7.
In this case, we cannot constrain configurations of models\footnote{According to Ref.\cite{Abe:2015_Classification}, the number of Higgs field was obtained for more than 5 and less than 9. We do not understand why the discrepancy between Ref.\cite{Abe:2015_Classification} and ours happened and do not discuss it further.}.

\subsubsection*{Case II : $\chi^{\text{\rm{L}}}=-\chi^{\text{\rm{R}}}$}
In this case, 4d left- and right-handed fermions come from the 6d Weyl fermion with different chirality. 
Eq.(\ref{constraint-2}) is written as
\begin{equation}
\forall I \in \{1,2,3,4\}, \qquad
-\rho^{\text{L}}_I + \rho^{\text{R}}_I+\rho^{\text{H}}_I=0 \mod 2.
\label{3-15-0}
\end{equation}
Taking $\rho^{\text{L,R,H}}_I$ to be 0 or 1, we have
\begin{equation}
|-\rho_I^{\text{L}} +\rho_I^{\text{R}}| = \rho_I^{\text{H}},
\label{3-15}
\end{equation}
therefore,
\begin{equation}
V^{\text{H}}= \sum_{I=1}^4|-\rho_I^{\text{L}} +\rho_I^{\text{R}}| .
\label{3-16}
\end{equation}
Since the left- and the right-handed quarks/leptons have opposite 6d chiralities each other, 
their 2d chiralities of zero modes are same. 
This means that $M^{\text{L}}$ and $M^{\text{R}}$ have same signs. 
For instance, if $\chi^{\text{L}}=-1$ and $\chi^{\text{R}}=+1$, 
$M^{\text{L}}$ and $M^{\text{R}}$ are positive definite because both zero modes come from 2d chirality $+1$. 
From the relation (\ref{constraint-2-1}),
\begin{equation}
M^{\text{H}} = M^{\text{L}} - M^{\text{R}} =
\begin{cases}
 |M^{\text{L}}| -|M^{\text{R}}|&\text{for }\chi^{\text{L}}=-1\text{ and }\chi^{\text{R}}=+1, \\
 -|M^{\text{L}}|+|M^{\text{R}}|&\text{for }\chi^{\text{L}}=+1\text{ and }\chi^{\text{R}}=-1,
\end{cases}
\label{3-16-0}
\end{equation}
hence we see $|M^{\text{H}}|=\Bigl||M^{\text{L}}|-|M^{\text{R}}|\Bigr|$ in either case. 
Now we would like to count the number of Higgs fields similar to Eq.(\ref{3-12}).
We note that flux quanta of Higgs field is
\begin{align}
|M^{\text{H}}|&=\Bigl||M^{\text{L}}|-|M^{\text{R}}|\Bigr| \nonumber \\
&=|V^{\text{L}}-V^{\text{R}}| \nonumber \\
&=\left|\sum_{I=1}^4 ( -\rho^{\text{L}}_I + \rho^{\text{R}}_I)\right|,
\label{3-16-1}
\end{align}
where we use Eq.(\ref{constraint-1}) in the second equality. 
Substituting Eqs.(\ref{3-16}) and (\ref{3-16-1}) into the condition (\ref{3-9}), 
we obtain
\begin{equation}
\left|\sum_{I=1}^4 ( -\rho^{\text{L}}_I + \rho^{\text{R}}_I)\right|  
\geq \sum_{I=1}^4|-\rho_I^{\text{L}} +\rho_I^{\text{R}}|.
\label{3-17}
\end{equation}
On the other hand, we know 
\begin{equation}
\left|\sum_{I=1}^4 ( -\rho^{\text{L}}_I + \rho^{\text{R}}_I)\right|  
\leq \sum_{I=1}^4|-\rho_I^{\text{L}} +\rho_I^{\text{R}}|
\label{3-18}
\end{equation}
from the triangle inequality. 
In order to hold both inequalities, we conclude
\begin{equation}
\left|\sum_{I=1}^4 ( -\rho^{\text{L}}_I + \rho^{\text{R}}_I)\right| 
= \sum_{I=1}^4|-\rho_I^{\text{L}} +\rho_I^{\text{R}}|.
\label{3-19}
\end{equation}
Eq.(\ref{3-19}) is satisfied if and only if $\Delta \rho_I \coloneqq -\rho_I^{\text{L}}+\rho_I^{\text{R}} \, (I=1,2,3,4)$ do not have different signs.

\begin{figure}[h]
    \centering
    \includegraphics[height=5cm]{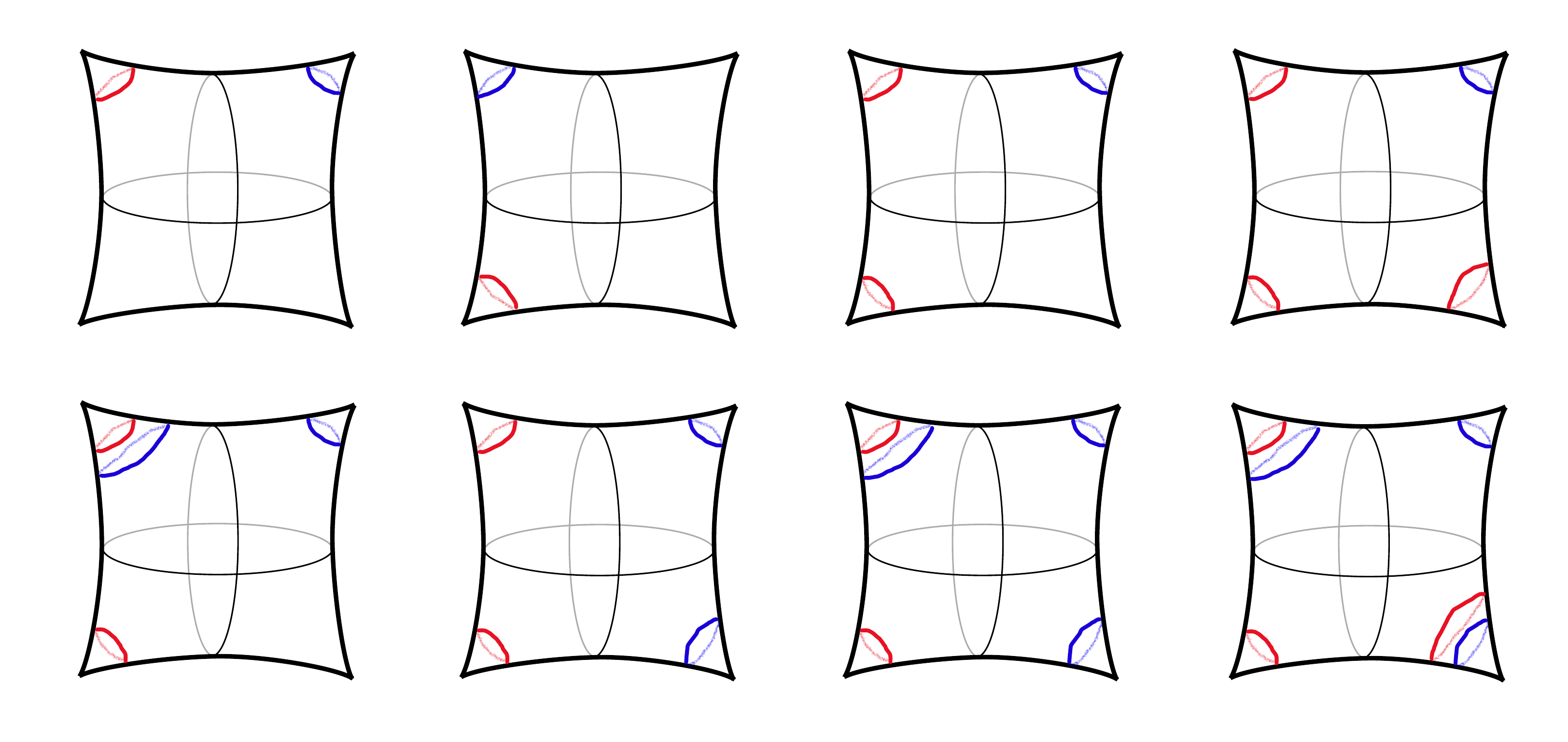}
    \caption{Examples of the forbidden configurations. The red(blue) lines show the windings of left-(right-) handed fermions. It is forbidden if there coexist fixed points with only the left-handed winding and those with only the right-handed winding.
     }
    \label{fig:3-A}%
\end{figure}

\begin{figure}[h]
    \centering
    \includegraphics[height=3cm]{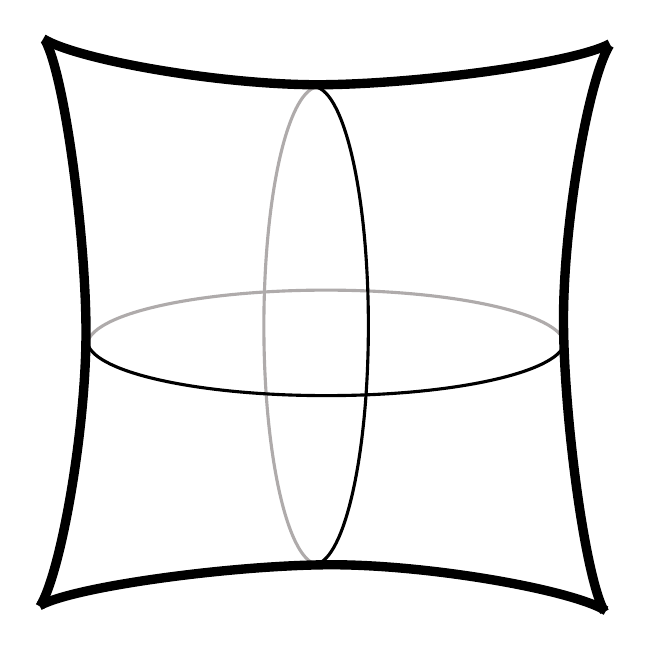}
    \caption{An example of the allowed configurations, where there is no winding at any fixed point.}
    \label{fig:3-B}%
\end{figure}

\begin{figure}[h]
    \centering
    \includegraphics[height=5cm]{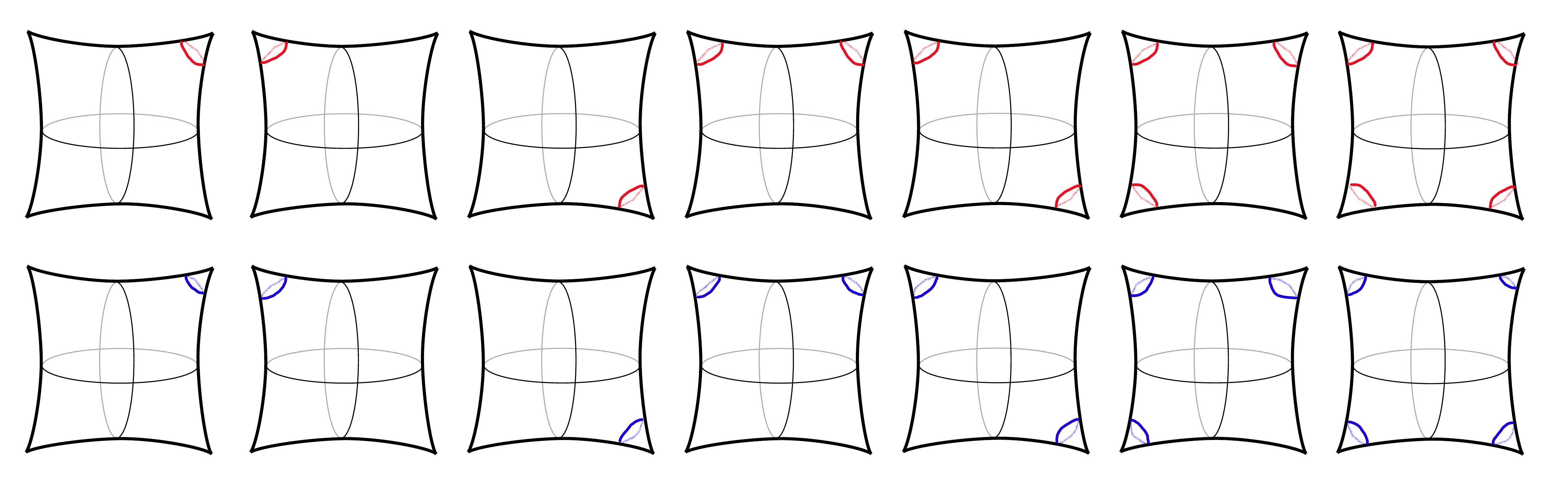}
    \caption{Examples of the allowed configurations, where there are only windings of left-handed fermion (upper) or right-handed fermion (lower).}
    \label{fig:3-C}%
\end{figure}

\begin{figure}[h]
    \centering
    \includegraphics[height=3cm]{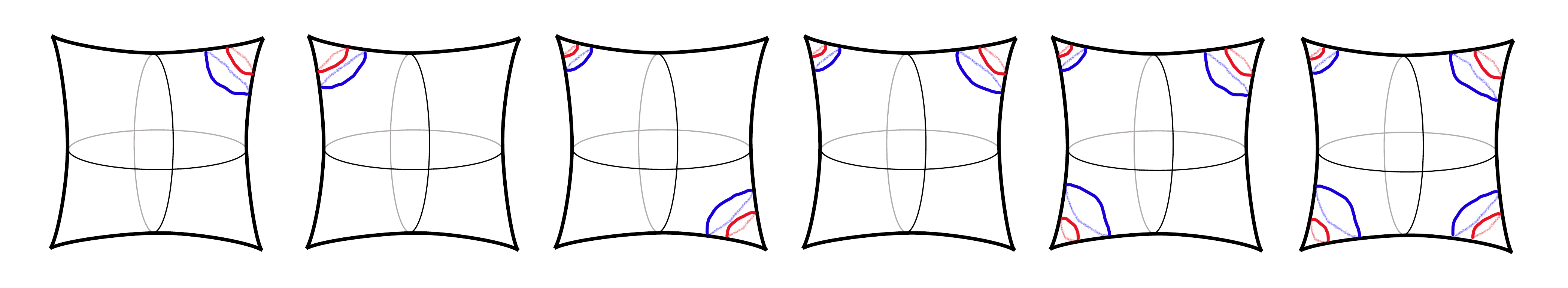}
    \caption{Examples of the allowed configurations, where both left- and right-handed fermions are winding at the same fixed points.}
    \label{fig:3-D}%
\end{figure}

\begin{figure}[h]
    \centering
    \includegraphics[height=3cm]{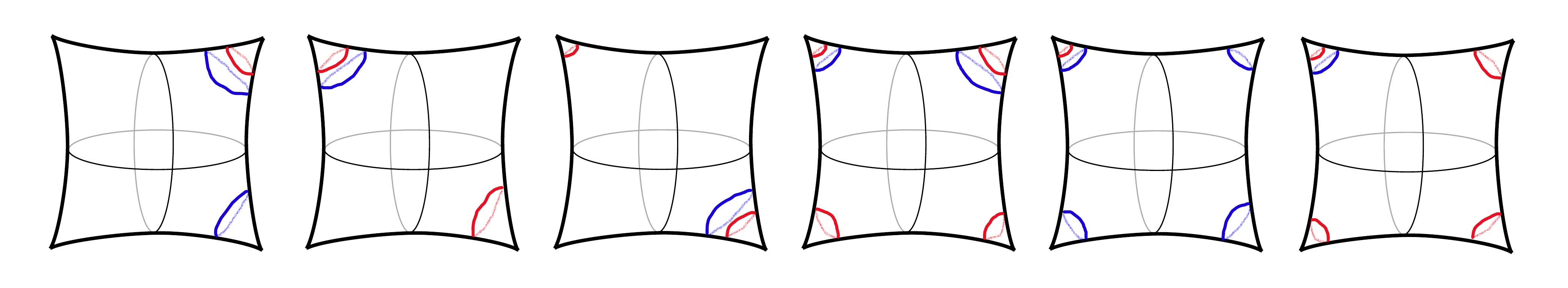}
    \caption{Examples of the allowed configurations. These are combinations of Fig.\ref{fig:3-C} and Fig.\ref{fig:3-D}.}
    \label{fig:3-E}%
\end{figure}

It is clear from a geometrical viewpoint what kind of configurations satisfy Eq.(\ref{3-19}). The configurations are forbidden if there coexist a fixed point where only the left-handed fermion is winding and it where only the right-handed fermion is winding (See Fig.\ref{fig:3-A}). The others are allowed (See Fig.\ref{fig:3-B} - \ref{fig:3-E}).

Let us see why the configurations in Fig.\ref{fig:3-A} are forbidden. If only the left-handed fermion is winding at a fixed point $z^{\text{fp}}_I$, $\Delta\rho_I =-1$.
On the other hand, if only the right-handed fermion is winding at a fixed point $z^{\text{fp}}_J$, $\Delta\rho_J=+1$. In the case of $I\neq J$, the left-hand side of Eq.(\ref{3-19}) becomes smaller since cancellation between $\Delta\rho_I$ and $\Delta\rho_J$ occurs.
Let us explain with an example of the upper leftmost picture of Fig.\ref{fig:3-A}, which is $\Delta\rho_1=\Delta\rho_2=0$, $\Delta \rho_3=-1$, $\Delta \rho_4=+1$. The left-hand side of (\ref{3-19}) is zero since $\Delta \rho_3$ and $\Delta \rho_4$ are cancelled each other, while the right-hand side of (\ref{3-19}) is two.

The other configurations in Fig.\ref{fig:3-B} - \ref{fig:3-E} are consistent with Eq.(\ref{3-19}).  The configuration with no any winding, which is illustrated in Fig.\ref{fig:3-B}, makes both sides zero. Configurations with windings of only left-(right-)handed fermion, which is illustrated in Fig.\ref{fig:3-C}, make Eq.(\ref{3-19}) hold due to $\rho_I^{\text{L}}=0(\rho_I^{\text{R}}=0)$ for any $I=1,2,3,4$. Fig.\ref{fig:3-D} shows that configuration whose fixed points are $\Delta\rho_I=0$, therefore both sides of Eq.(\ref{3-19}) are zero. Configurations in Fig.\ref{fig:3-E} have both the properties of Fig.\ref{fig:3-C} and Fig.\ref{fig:3-D}, whose fixed points are $\Delta\rho_I=0$ otherwise $\rho_I^{\text{L}}=0$ or $\rho_I^{\text{R}}=0$.

Through the above analysis, we see that allowed and forbidden configurations are classified geometrically from pictures. Then we can predict Higgs numbers.
Applying Eq.(\ref{3-19}) to Eq.(\ref{3-8}), we find $n^{\text{H}} =1$ independent of the generation number $g$.
Thus, the realistic models in this case have only one Higgs fileds.
We also count allowed configurations in Appendix \ref{Appendix-2} and the result is consistent with Table 16 in Ref.\cite{Abe:2015_Classification}.
We emphasize that the constraint on the models is given by winding numbers, which are geometrically clear from figures.



\section{Conclusion and discussion}
\label{sec:4}
In this paper, we have investigated how the conditions on generation numbers and Yukawa coupling can constrain models with $T^2/\mathbb{Z}_2$ flux compactification. As a result, we found that there is a geometry behind realistic models in a 6d chirality assignment. The geometry which governs phenomenology is given by winding numbers at the orbifold fixed points.

In the case that the 4d left- and right-handed fermions are originated from the same 6d Weyl fermion, we could not exclude any configurations by use of winding numbers. On the other hand, in the case that the 4d left- and right-handed fermions are originated from the 6d Weyl fermion with different chirality, the configurations are forbidden, where there is a fixed point with only the left-handed fermion's winding and another fixed point with only the right-handed fermion's winding (See Fig.\ref{fig:3-A}).

There are three remarkable points in this study by use of winding numbers at the fixed points. The first one is that we succeeded in illustrating realistic and unrealistic configurations of models. We summarize the forbidden configurations in Table \ref{table:2} - \ref{table:4} in Appendix \ref{Appendix-2}, but it seems hard to see the property of models intuitively. It is clear only if we illustrate as shown in Fig.\ref{fig:3-A}. Although we show allowed configurations in Fig.\ref{fig:3-B} - \ref{fig:3-E}, not all of these are allowed. More phenomenological constraints such as realization of mass hierarchies are desired to be studied.

The second one is that we do not need complicated calculations in our analysis. Thanks to winding numbers, we only need the triangle inequality (\ref{3-18}) to constrain the models in Case II of Subsection \ref{subsec:3-3}. The calculation in this paper is more straightforward and clearly than previous works such as Ref.\cite{Abe:2015_Classification}. 

The last one is that we do not have to assume three-generation of fermions. Therefore, our analysis is applicable if the fourth generation of quark/lepton are discovered in future experiments. The results were made possible thanks to winding numbers and generation number counting fourmula (\ref{2-27}). Without these, we had to specify generation number of fermions because of technical reasons in earlier works such as Refs.\cite{Abe:2015_Classification, Fujimoto:2016}. We expect an extension to more general cases of these work by use of winding numbers.

Finally, we comment on future works. It would be interesting to apply our analysis to models in other orbifold compactifications $T^2/\mathbb{Z}_N\, (N=3,4,6)$, where winding numbers have been studied in Refs.\cite{Sakamoto:2020, Imai:2022, Sakamoto:2021}. We hope that our analysis would be further powerful to find realistic models in such cases.


\appendix
\renewcommand{\thesection}{\Alph{section}}

\section{Check of constraints (\ref{constraint-2})}
\label{Appendix-1}
In this Appendix, we show that the four conditions (\ref{constraint-2-1})-(\ref{constraint-2-3}) and (\ref{constraint-2-4}) are rewritten as Eq.(\ref{constraint-2}). At first, let us check 
\begin{equation}
-\chi^{\text{L}}\chi^{\text{R}}  = (-1)^{(\chi^{\text{L}}+\chi^{\text{R}} )/2}.
\label{proof-0}
\end{equation}
If $\chi^{\text{L}}=\chi^{\text{R}}$, $\chi^{\text{L}}\chi^{\text{R}} =1$ and $(\chi^{\text{L}}+\chi^{\text{R}} )/2=\pm1$ holds, therefore both sides of Eq.(\ref{proof-0}) are $-1$. On the other hand if $\chi^{\text{L}}=-\chi^{\text{R}}$, $\chi^{\text{L}}\chi^{\text{R}} =-1$ and $(\chi^{\text{L}}+\chi^{\text{R}} )/2=0$ holds, therefore both sides of Eq.(\ref{proof-0}) are $1$. In either case, we Eq.(\ref{proof-0}) holds.

\begin{itemize}
\item[$I=1$:]At $z_1^{\text{fp}}=0$, the winding number $\rho_1^f$ is defined from Eq.(\ref{2-25}) as
\begin{equation}
\lim_{z\to 0}\psi^f_{T^2/\mathbb{Z}_2,\pm, n}(-z)_{\eta^f} = (-1)^{\rho_1^f}\lim_{z\to 0}\psi^f_{T^2/\mathbb{Z}_2,\pm, n}(z)_{\eta^f}.
\label{proof-1}
\end{equation}
Eq.(\ref{proof-1}) is nothing but the definition of the $\mathbb{Z}_2$ eigenvalue $\eta^f$.
From Eqs.(\ref{constraint-2-4}) with (\ref{proof-0}), we obtain
\begin{equation}
(-1)^{(\chi^{\text{L}} + \chi^{\text{R}})/2} \,  (-1)^{-\rho_1^{\text{L}}} (-1)^{\rho_1^{\text{H}}}(-1)^{\rho_1^\text{H}} = 1, \label{proof-2}
\end{equation}
therefore Eq.(\ref{constraint-2}) is verified in $I=1$.
\item[$I=2$:]At $z_2^{\text{fp}}=1/2$, it is a bit nontrivial. The definition of $\rho_2^f$ is given by
\begin{equation}
\lim_{z\to 0}\psi^f_{T^2/\mathbb{Z}_2,\pm, n}\!\left(\frac12-z\right)_{\! \! \eta^f} \!\!\! = (-1)^{\rho_2^f}\lim_{z\to 0}\psi^f_{T^2/\mathbb{Z}_2,\pm, n}\!\left(\frac12+z\right)_{\! \! \eta^f}.
\label{proof-3}
\end{equation}
The left-hand side of Eq.(\ref{proof-3}) is deformed as
\begin{align}
\lim_{z\to 0}\psi^f_{T^2/\mathbb{Z}_2,\pm, n}\!\left(\frac12-z\right)_{\! \! \eta^f} \!\!\! 
&=\lim_{z\to 0} e^{iq_f \Lambda_1\left(-\frac12 -z\right) +2i\pi \alpha_1^f}\psi^f_{T^2/\mathbb{Z}_2,\pm, n}\!\left(-\frac12-z\right)_{\! \! \eta^f}  \nonumber \\
&=\lim_{z\to 0} e^{iq_f \Lambda_1\left(-\frac12 -z\right) +2i\pi \alpha_1^f}\eta^f \psi^f_{T^2/\mathbb{Z}_2,\pm, n}\!\left(\frac12+z\right)_{\! \! \eta^f} \nonumber \\
&=\eta^f e^{2i\pi \alpha_1^f}\lim_{z\to 0}\psi^f_{T^2/\mathbb{Z}_2,\pm, n}\!\left(\frac12+z\right)_{\! \! \eta^f},
\label{proof-4}
\end{align}
where we use the torus conditon (\ref{2-10-1}) in the first equality,
and $\Lambda_1(-1/2)=0$ from (\ref{2-6}) in the last equality. Now, we find
\begin{equation}
(-1)^{\rho_2^f}=\eta^f e^{2i\pi\alpha_1^f}
\label{proof-5}
\end{equation}
by comparing Eq.(\ref{proof-3}) with Eq.(\ref{proof-4}).
From Eq.(\ref{constraint-2-2}) and Eq.(\ref{constraint-2-4}) with (\ref{proof-0}), we obtain
\begin{equation}
(-1)^{(\chi^{\text{L}} + \chi^{\text{R}})/2} \,  (-1)^{-\rho_2^{\text{L}}} (-1)^{\rho_2^{\text{H}}}(-1)^{\rho_2^\text{H}} =(-1)^{(\chi^{\text{L}} + \chi^{\text{R}})/2} \,  \bar\eta^{\text{L}} \eta^{\text{H}}\eta^{\text{R}} e^{2i\pi(-\alpha_1^{\text{L}}+\alpha_1^{\text{H}}+\alpha_1^{\text{R}})}=1.
\label{proof-6}
\end{equation}
\item[$I=3$:]At $z_3^{\text{fp}}=i/2$, we can verify as same as $I=2$. The definition of $\rho^f_3$ is given by
\begin{equation}
\lim_{z\to 0}\psi^f_{T^2/\mathbb{Z}_2,\pm, n}\!\left(\frac{i}2-z\right)_{\! \! \eta^f} \!\!\! = (-1)^{\rho_3^f}\lim_{z\to 0}\psi^f_{T^2/\mathbb{Z}_2,\pm, n}\!\left(\frac{i}2+z\right)_{\! \! \eta^f}.
\label{proof-7}
\end{equation}
The left-hand side of Eq.(\ref{proof-7}) is deformed as
\begin{align}
\lim_{z\to 0}\psi^f_{T^2/\mathbb{Z}_2,\pm, n}\!\left(\frac{i}2-z\right)_{\! \! \eta^f} \!\!\!
&=\lim_{z\to 0} e^{iq_f \Lambda_2\left(-\frac i2 -z\right) +2i\pi \alpha_2^f}\psi^f_{T^2/\mathbb{Z}_2,\pm, n}\!\left(-\frac{i}2-z\right)_{\! \! \eta^f} \nonumber \\
&=\lim_{z\to 0} e^{iq_f \Lambda_2\left(-\frac i2 -z\right) +2i\pi \alpha_2^f}\eta^f \psi^f_{T^2/\mathbb{Z}_2,\pm, n}\!\left(\frac{i}2+z\right)_{\! \! \eta^f} \nonumber \\
&=\eta^f e^{2i\pi \alpha_2^f}\lim_{z\to 0}\psi^f_{T^2/\mathbb{Z}_2,\pm, n}\!\left(\frac{i}2+z\right)_{\! \! \eta^f},
\label{proof-8}
\end{align}
where we use the torus conditon (\ref{2-10-2}) in the first equality,
and $\Lambda_2(-i/2)=0$ from Eq.(\ref{2-6}) in the last equality. We find
\begin{equation}
(-1)^{\rho_3^f}=\eta^f e^{2i\pi\alpha_2^f}
\label{proof-9}
\end{equation}
by comparing Eq.(\ref{proof-7}) with Eq.(\ref{proof-8}).
From Eq.(\ref{constraint-2-3}) and Eq.(\ref{constraint-2-4}), we obtain
\begin{equation}
(-1)^{(\chi^{\text{L}} + \chi^{\text{R}})/2} \,  (-1)^{-\rho_3^{\text{L}}} (-1)^{\rho_3^{\text{H}}}(-1)^{\rho_3^\text{H}} =(-1)^{(\chi^{\text{L}} + \chi^{\text{R}})/2} \,  \bar\eta^{\text{L}} \eta^{\text{H}}\eta^{\text{R}} e^{2i\pi(-\alpha_2^{\text{L}}+\alpha_2^{\text{H}}+\alpha_2^{\text{R}})}=1.
\label{proof-10}
\end{equation}
\item[$I=4$:]Finally, we show at $z_4^{\text{fp}}=(1+i)/2$. The definition of $\rho^f_4$ is given by
\begin{equation}
\lim_{z\to 0}\psi^f_{T^2/\mathbb{Z}_2,\pm, n}\!\left(\frac{1+i}2-z\right)_{\! \! \eta^f} \!\!\!= (-1)^{\rho_4^f}\lim_{z\to 0}\psi^f_{T^2/\mathbb{Z}_2,\pm, n}\!\left(\frac{1+i}2+z\right)_{\! \! \eta^f} .
\label{proof-11}
\end{equation}
The left-hand side of Eq.(\ref{proof-11}) is deformed as
\begin{align}
&\lim_{z\to 0}\psi^f_{T^2/\mathbb{Z}_2,\pm, n}\!\left(\frac{1+i}2-z\right)_{\! \! \eta^f} \!\!\!
\nonumber \\
&=\lim_{z\to 0} e^{iq_f \Lambda_2\left(\frac{1-i}2 -z\right) +2i\pi \alpha_2^f}\psi^f_{T^2/\mathbb{Z}_2,\pm, n}\!\left(\frac{1-i}2-z\right)_{\! \! \eta^f}\nonumber \\
&=\lim_{z\to 0} e^{iq_f \Lambda_2\left(\frac{1-i}2 -z\right) +2i\pi \alpha_2^f}e^{iq_f \Lambda_1\left(-\frac{1+i}2 -z\right) +2i\pi \alpha_1^f} \psi^f_{T^2/\mathbb{Z}_2,\pm, n}\!\left(-\frac{1+i}2-z\right)_{\! \! \eta^f} \nonumber \\
&=\lim_{z\to 0} e^{iq_f \Lambda_2\left(\frac{1-i}2 -z\right) +2i\pi \alpha_2^f}e^{iq_f \Lambda_1\left(-\frac{1+i}2 -z\right) +2i\pi \alpha_1^f}\eta^f \psi^f_{T^2/\mathbb{Z}_2,\pm, n}\!\left(\frac{1+i}2+z\right)_{\! \! \eta^f} \nonumber \\
&= e^{-i\pi M^f+2i\pi \alpha_1^f+2i\pi \alpha_2^f}\eta^f \lim_{z\to 0} \psi^f_{T^2/\mathbb{Z}_2,\pm, n}\!\left(\frac{1+i}2+z\right)_{\! \! \eta^f},
\label{proof-12}
\end{align}
where we use the torus conditon (\ref{2-10-2}) and (\ref{2-10-1}) in the first and second equality, respectively. In the last equality, we use Eq.(\ref{2-6}) and flux quantized condition (\ref{2-quanta}). We find
\begin{equation}
(-1)^{\rho_4^f}=\eta^fe^{-i\pi M^f+2i\pi \alpha_1^f+2i\pi \alpha_2^f}
\label{proof-13}
\end{equation}
by comparing Eq.(\ref{proof-11}) with Eq.(\ref{proof-12}).
From Eq.(\ref{constraint-2-1})-(\ref{constraint-2-3}) and Eq.(\ref{constraint-2-4}), we obtain
\begin{align}
&(-1)^{(\chi^{\text{L}} + \chi^{\text{R}})/2} \,  (-1)^{-\rho_4^{\text{L}}} (-1)^{\rho_4^{\text{H}}}(-1)^{\rho_4^\text{H}} \nonumber \\
&=(-1)^{(\chi^{\text{L}} + \chi^{\text{R}})/2} \,  \bar\eta^{\text{L}} \eta^{\text{H}}\eta^{\text{R}} 
e^{-i\pi(-M^{\text{L}}+M^{\text{H}}+M^{\text{R}})}e^{2i\pi(-\alpha_1^{\text{L}}+\alpha_1^{\text{H}}+\alpha_1^{\text{R}})}e^{2i\pi(-\alpha_2^{\text{L}}+\alpha_2^{\text{H}}+\alpha_2^{\text{R}})}=1.
\label{proof-14}
\end{align}
\end{itemize}

\section{Allowed and forbidden configurations in Case II}
\label{Appendix-2}

We show all configurations of Case II in \ref{subsec:3-3} and analyze whether each configuration is allowed or not.

At first, we show that all cases with the classes $V^{\text{L}}=0$ and $V^{\text{L}}=4$ are allowed. 
The important feature in both classes is that all winding numbers $\rho_{I}^{\text{L}} \, (I=1,2,3,4)$ are same value. 
In the class $V^{\text{L}}=0$, 
the only pattern $(\rho^{\text{L}}_1,\rho^{\text{L}}_2,\rho^{\text{L}}_3,\rho^{\text{L}}_4)=(0,0,0,0)$ is allowed, 
thus this means $\Delta\rho_I = \rho^{\text{R}}_I$ for all $I\in\{1,2,3,4\}$, which takes $0$ or $+1$. 
On the other hand, in the class $V^{\text{L}}=4$,  
the only pattern $(\rho^{\text{L}}_1,\rho^{\text{L}}_2,\rho^{\text{L}}_3,\rho^{\text{L}}_4)=(1,1,1,1)$ is allowed, 
 thus this implies $\Delta \rho_I = -1+\rho^{\text{R}}_I$ for all $I\in\{1,2,3,4\}$, which takes $0$ or $-1$. 
In either case, since all of $\Delta\rho_I \, (I=1,2,3,4)$ have same signs, 
the patterns in the left-handed classes $V^{\text{L}}=0,4$ satisfy (\ref{3-19}). 
Replacing $V^{\text{L}}$ with $V^{\text{R}}$, the similar argument holds for $V^{\text{R}}$, 
i.e. any patterns with the right-handed classes $V^{\text{R}}=0,4$ are allowed. 

What we should discuss carefully are patterns of winding numbers 
 in the cases of $V^{\text{L}}=1,2,3$ and $V^{\text{R}}=1,2,3$. 
We will find some forbidden patterns and analyze for each left-handed class.

\subsubsection*{\underline{$V^{\rm{L}}=1$}}
In this case, the allowed patterns are that one of the winding numbers 
$\rho^{\text{L}}_I \, (I=1,2,3,4)$ is $1$ and the others are $0$. 
It is convenient to rewrite (\ref{3-19}) as
\begin{equation}
|-1+V^{\text{R}}| = |-1+\rho^{\text{R}}_I|+\sum_{J\neq I}|\rho_J^{\text{R}}|.
\label{3-19-0}
\end{equation}
Eq.(\ref{3-19-0}) holds if and only if $\rho^{\text{R}}_I=1$. 
In order to understand this, let us study an example $(\rho_1^{\text{L}},\rho_2^{\text{L}},\rho_3^{\text{L}},\rho_4^{\text{L}})=(1,0,0,0)$ for each right-handed class $V^{\text{R}}=1,2,3$. Eq.(\ref{3-19-0}) is written as
\begin{equation}
|-1+V^{\text{R}}| = |-1+\rho^{\text{R}}_1|+\sum_{J=2}^4|\rho_J^{\text{R}}|.
\label{3-19-1}
\end{equation}

\begin{itemize}
\item For $V^{\text{R}}=1$, the left-hand side of (\ref{3-19-1}) is 0. 
$V^{\text{R}}=1$ also means that one of $\rho^{\text{R}}_I \, (I=1,2,3,4)$ are 1 and the others are 0. 
Therefore, $\rho^{\text{R}}_1=1$ and $\rho^{\text{R}}_J=0 \, (J=2,3,4)$ are required to satisfy (\ref{3-19-1}). 
That is, only $(\rho_1^{\text{R}},\rho_2^{\text{R}},\rho_3^{\text{R}},\rho_4^{\text{R}})=(1,0,0,0)$ is allowed 
in the class $V^{\text{L}}=1$.

\item For $V^{\text{R}}=2$, the left-hand side of (\ref{3-19-1}) is 1. 
$V^{\text{R}}=2$ also means that two of $\rho^{\text{R}}_I \, (I=1,2,3,4)$ are 1 and the other two are 0. 
If we take $\rho^{\text{R}}_1=0$, the first term of (\ref{3-19-1}) is $1$, and the second term becomes 2, 
thus these patterns are forbidden. 
From this observation, the following three patterns are found to be allowed: 
$(\rho_1^{\text{R}},\rho_2^{\text{R}},\rho_3^{\text{R}},\rho_4^{\text{R}})=(1,0,0,1), (1,0,1,0),(1,1,0,0)$. 

\item For $V^{\text{R}}=3$, the left-hand side of (\ref{3-19-1}) is 2. 
$V^{\text{R}}=3$ also means that three of $\rho^{\text{R}}_I \, (I=1,2,3,4)$ are 1 and the remaining one is 0. 
If we take $\rho^{\text{R}}_1=0$, the first term of (\ref{3-19-1}) is $1$, 
and the second term becomes 3, thus these patterns are forbidden. 
The other three patterns are found to be allowed: 
$(\rho_1^{\text{R}},\rho_2^{\text{R}},\rho_3^{\text{R}},\rho_4^{\text{R}})=(1,0,1,1), (1,1,0,1),(1,1,1,0)$.
\end{itemize}
We summarize our results of allowed patterns of winding numbers in the case of $V^{\text{L}}=1$. 
The patterns except for $\times$ shown in Table \ref{table:2} are allowed. 
There are nine allowed patterns for each left-handed pattern. 
In total, 36 configurations are allowed for the class $V^{\text{L}}=1$.

\begin{table}[h]
\centering
	\begin{tabular}{c|c|cccc}
	\hline \hline
	\addlinespace[2pt]
	$V^{\text{R}}$&$(\rho_1^{\text{R}},\rho_2^{\text{R}},\rho_3^{\text{R}},\rho_4^{\text{R}})$
	& (1,0,0,0)&(0,1,0,0)&(0,0,1,0)&(0,0,0,1) \\
	\hline
	0&(0,0,0,0)&&&& \\
	\hline
	\multirow{4}{*}{1}&(0,0,0,1)&$\times$&$\times$&$\times$&\\
	&(0,0,1,0)&$\times$&$\times$&&$\times$ \\
	&(0,1,0,0)&$\times$&&$\times$&$\times$  \\
	&(1,0,0,0)&&$\times$&$\times$&$\times$ \\
	\hline
	\multirow{6}{*}{2}&(0,0,1,1)&$\times$&$\times$&&\\
	&(0,1,0,1)&$\times$&&$\times$& \\
	&(0,1,1,0)&$\times$&&&$\times$ \\
	&(1,0,0,1)&&$\times$&$\times$& \\
	&(1,0,1,0)&&$\times$&&$\times$ \\
	&(1,1,0,0)&&&$\times$&$\times$ \\
	\hline
	\multirow{4}{*}{3}&(0,1,1,1)&$\times$&&&\\
	&(1,0,1,1)&&$\times$&& \\
	&(1,1,0,1)&&&$\times$& \\
	&(1,1,1,0)&&&&$\times$ \\
	\hline
	4&(1,1,1,1)&&&& \\
	\hline \hline
	\end{tabular}
	\caption{Forbidden patterns in the class of $V^{\text{L}}=1$. 
	The symbol $\times$ represents that the pattern is forbidden. 
	In the top right row, the possible combinations of $V^{\text{L}}=1$ are listed.}
	\label{table:2}
\end{table}

\subsubsection*{\underline{$V^{\rm{L}}=2$}}
Two of four winding numbers for the left-handed fermions are 1 and the remaining two are 0. 
Setting $\rho_{I_1}^{\text{L}}=\rho_{I_2}^{\text{L}}=1$ and $\rho_{I_3}^{\text{L}}=\rho_{I_3}^{\text{L}}=0$, 
where $I_1,I_2,I_3,I_4\in\{1,2,3,4\}$, 
we have $\Delta\rho_{I_1}=-1+\rho^{\text{R}}_{I_1}$, $\Delta\rho_{I_2}=-1+\rho^{\text{R}}_{I_2}$, 
$\Delta\rho_{I_3}=\rho^{\text{R}}_{I_3}$, $\Delta\rho_{I_4}=\rho^{\text{R}}_{I_4}$. 
In this case, (\ref{3-19}) becomes as follows. 
\begin{align}
\left|-2 + V^{\text{R}}_I \right| 
= |-1 +\rho_{I_1}^{\text{R}}| + |-1 +\rho_{I_2}^{\text{R}}|
+ |\rho_{I_3}^{\text{R}}| + |\rho_{I_4}^{\text{R}}|.
\label{3-19-2}
\end{align}

\begin{itemize}
\item For $V^{\text{R}}=1$ case, the solution of winding numbers to (\ref{3-19-2}) are found as
$(\rho_{I_1}^{\text{R}},\rho_{I_2}^{\text{R}},\rho_{I_3}^{\text{R}}, \rho_{I_4}^{\text{R}})=(1, 0, 0, 0), (0, 1, 0, 0)$.
\item For $V^{\text{R}}=2$ case, the solution of winding numbers to (\ref{3-19-2}) is found as 
$(\rho_{I_1}^{\text{R}},\rho_{I_2}^{\text{R}},\rho_{I_3}^{\text{R}}, \rho_{I_4}^{\text{R}})=(1, 1, 0, 0)$.
\item For $V^{\text{R}}=3$ case, the solution of winding numbers to (\ref{3-19-2}) are found as
$(\rho_{I_1}^{\text{R}},\rho_{I_2}^{\text{R}},\rho_{I_3}^{\text{R}}, \rho_{I_4}^{\text{R}})=(1, 1, 1, 0), (1, 1, 0, 1)$.
\end{itemize}
As a result, the patterns of the winding numbers except for $\times$ shown in Table \ref{table:3} are allowed. 
There are seven allowed patterns for each left-handed pattern. 
In total, 42 patterns are allowed for the class $V^{\text{L}}=2$.
\begin{table}[h]
\centering
	\begin{tabular}{c|c|cccccc}
	\hline \hline
	\addlinespace[2pt]
	$V^{\text{R}}$&$(\rho_1^{\text{R}},\rho_2^{\text{R}},\rho_3^{\text{R}},\rho_4^{\text{R}})$& (0,0,1,1)&(0,1,0,1)&(1,0,0,1)&(0,1,1,0)&(1,0,1,0)&(1,1,0,0) \\
	\hline
	0&(0,0,0,0)&&&&&& \\
	\hline
	\multirow{4}{*}{1}&(0,0,0,1)&&&& $\times$ & $\times$ & $\times$ \\
	&(0,0,1,0)&& $\times$ &$\times$&&& $\times$ \\
	&(0,1,0,0)& $\times$ &&$\times$&& $\times$ & \\
	&(1,0,0,0)& $\times$ & $\times$ &&$\times$&& \\
	\hline
	\multirow{6}{*}{2}&(0,0,1,1)&&$\times$&$\times$&$\times$&$\times$& $\times$ \\
	&(0,1,0,1)&$\times$&&$\times$&$\times$&$\times$&$\times$ \\
	&(0,1,1,0)&$\times$&$\times$&$\times$&&$\times$&$\times$ \\
	&(1,0,0,1)&$\times$&$\times$&&$\times$&$\times$&$\times$ \\
	&(1,0,1,0)&$\times$&$\times$&$\times$&$\times$&&$\times$ \\
	&(1,1,0,0)&$\times$&$\times$&$\times$&$\times$&$\times$& \\
	\hline
	\multirow{4}{*}{3}&(0,1,1,1)&&&$\times$&&$\times$&$\times$\\
	&(1,0,1,1)&&$\times$&&$\times$&&$\times$ \\
	&(1,1,0,1)&$\times$&&&$\times$&$\times$& \\
	&(1,1,1,0)&$\times$&$\times$&$\times$&&& \\
	\hline
	4&(1,1,1,1)&&&& \\
	\hline \hline
	\end{tabular}
	\caption{Forbidden patterns in the class of $V^{\text{L}}=2$. 
	The symbol $\times$ represents that the pattern is forbidden. 
	In the top right row, the possible combinations of $V^{\text{L}}=2$ are listed.}
	\label{table:3}
\end{table}
\subsubsection*{\underline{$V^{\rm{L}}=3$}}
We can understand in the case of $V^{\text{L}}=3$ as same as $V^{\text{L}}=1$, 
because one of the winding numbers $\rho^{\text{L}}_I \, (I=1,2,3,4)$ is 0 and others are 1. 
In this case, (\ref{3-19}) becomes 
\begin{align}
\left|-3 + V^{\text{R}}_I \right| 
= |\rho_{I}^{\text{R}}| + \sum_{J \ne I}^4|-1 + \rho_J^{\text{R}}|
\label{3-19-3}
\end{align}
and $\rho^{\text{R}}_I=0$ is required if $V^{\text{R}}=1,2,3$.
The patterns of the winding numbers except for $\times$ shown in Table \ref{table:4} are allowed. 
There are nine allowed patterns for each left-handed pattern. 
In total, 36 configurations are allowed for the class $V^{\text{L}}=3$.
\begin{table}[h]
\centering
	\begin{tabular}{c|c|cccc}
	\hline \hline
	\addlinespace[2pt]
	$V^{\text{R}}$&$(\rho_1^{\text{R}},\rho_2^{\text{R}},\rho_3^{\text{R}},\rho_4^{\text{R}})$& (1,1,1,0)&(1,1,0,1)&(1,0,1,1)&(0,1,1,1) \\
	\hline
	0&(0,0,0,0)&&&& \\
	\hline
	\multirow{4}{*}{1}&(0,0,0,1)&$\times$&&&\\
	&(0,0,1,0)&&$\times$&& \\
	&(0,1,0,0)&&&$\times$&  \\
	&(1,0,0,0)&&&&$\times$ \\
	\hline
	\multirow{6}{*}{2}&(0,0,1,1)&$\times$&$\times$&&\\
	&(0,1,0,1)&$\times$&&$\times$& \\
	&(0,1,1,0)&&$\times$&$\times$& \\
	&(1,0,0,1)&$\times$&&&$\times$ \\
	&(1,0,1,0)&&$\times$&&$\times$ \\
	&(1,1,0,0)&&&$\times$&$\times$ \\
	\hline
	\multirow{4}{*}{3}&(0,1,1,1)&$\times$&$\times$&$\times$&\\
	&(1,0,1,1)&$\times$&$\times$&&$\times$ \\
	&(1,1,0,1)&$\times$&&$\times$&$\times$ \\
	&(1,1,1,0)&&$\times$&$\times$&$\times$ \\
	\hline
	4&(1,1,1,1)&&&& \\
	\hline \hline
	\end{tabular}
	\caption{Forbidden patterns in the class of $V^{\text{L}}=3$. 
	The symbol $\times$ represents that the pattern is forbidden. 
	In the top right row, the possible combinations of $V^{\text{L}}=3$ are listed.}
	\label{table:4}
\end{table}

In summary, 146 patterns of the total number 
of possible patterns of winding numbers $16\times16=256$ are allowed. 
They are classified 16 patterns with $V^{\text{L}}=0$, $4\times 9=36$ patterns with $V^{\text{L}}=1$, 
$6\times 7=42$ patterns with $V^{\text{L}}=2$, $4\times 9=36$ patterns with $V^{\text{L}}=3$, 
16 patterns with $V^{\text{L}}=4$. 

\bibliographystyle{utphys}
 \bibliography{1}

\end{document}